**The following manuscript is accepted for publication in Journal of Materials Chemistry, DOI:10.1039/C2JM30251A. Please use this new link.**

Compared with the old version, we revised the errors in it. We also tried to get the ratio of $Bi_8(AlCl_4)_2$ to $Bi_5(AlCl_4)_3$, based on the suggestions from the co-authors and colleagues, but did not succeed, possibly due to the poor signal to noise ratio of XRD spectrum.

There are one simple route for further obtaining detailed information of $Bi_8^{2+}$.

One can synthesize high-quality single crystal through method developed in Eur. J. Inorg. Chem., 2005, 670.

This result further suggests that the establishment of structure-property relationships is the first and the most important step to get a clearer picture of the PL mechanism related to bismuth species.

# Experimental and Theoretical Studies of Photoluminescence from $Bi_8^{2+}$ and $Bi_5^{3+}$ stabilized by $[AlCl_4]^-$ in Molecular Crystals

Hong-Tao Sun,*[a,b] Yoshio Sakka,[c] Naoto Shirahata,[c,d] Hong Gao,[a] and Tetsu Yonezawa[b]

The photophysical properties of $Bi_8^{2+}$ and $Bi_5^{3+}$ polycations stabilized by $[AlCl_4]^-$ have been studied experimentally and theoretically. The obtained product was thoroughly evaluated by powder X-ray diffraction and photoluminescence spectroscopy, making it clear that both $Bi_8^{2+}$ and $Bi_5^{3+}$ contribute to the observed broad near-infrared emission. Furthermore, it was revealed that $Bi_8^{2+}$ polycation mainly
10 results in the emission peaking at ca. 1180 nm, while $Bi_5^{3+}$ the longer-wavelength emission. The following quantum chemistry calculation on $Bi_8^{2+}$ polycation helps us attribute some observed excitation bands in the visible spectral range to specific electronic transitions of bismuth polycations. It is believed that systematical investigation of structural and luminescent properties as well as detailed quantum chemistry calculation of molecular crystals containing such kinds of bismuth units allows us to obtain a
15 clearer picture on bismuth-related photophysical behaviors, which not only serve to solve the confusions on the luminescence origin of bismuth in other material systems such as bulk glasses, glass fibers and conventional crystals, but also is helpful to develop novel applicable broadband tunable laser mediums.

## Introduction

Bismuth is one of the most investigated main group elements,
20 which has been called "the wonder metal" owing to its peculiar physiochemical properties.[1] It is well known that in addition to the most stable form of Bi (i.e., $Bi^{3+}$), subvalent Bi could exist in a wide range of material systems such as molten salts and molecular crystals, which demonstrates peculiar photophysical
25 properties.[1,2] In the past decade, bismuth activated near-infrared (NIR) emitting materials have received extensive attention owing to their peculiar features such as broad and long-lived photoluminescence (PL) band covering the telecommunication and optimal biological optical windows, high quantum efficiency,
30 and host-dependence spectral lineshape.[3-16] So far, NIR PL from Bi has been observed in diverse materials such as glasses,[3-4] conventional crystals (refs.5-11) and ionic liquids (ref.12), and functional applications for lasers and bioimaging were successfully demonstrated in 2005 (ref. 3b) and 2011 (ref. 4),
35 respectively. In contrast to above advances, the PL mechanisms from these systems remain hotly debated.[2-13] Clearly, there is not a convincing model for all materials, and the PL assignment should be correlated with the hosts used and practical preparation procedures. Indeed, the fundamental roadblock facing the PL
40 origin of Bi doped materials is the unknown local coordination environment of Bi in most hosts, although some attempts have been tried in Bi doped crystalline materials.[5,6a,7-11] This uncertainty directly results in many kinds of controversial assignments on the NIR PL from Bi. Given this issue, it is
45 obvious that not only the preparations but also structural analyses of optically active materials containing Bi are of vital importance in order to obtain a convincing picture on the PL origin. Recently, Sun *et al.* reported that $Bi_5(AlCl_4)_3$ molecular crystal exhibits extremely broad NIR PL with a full width at the half maximum
50 (FWHM) of > 510 nm.[14] A major difference favoring this crystal over conventional Bi doped materials is its clear crystalline structure, i.e., well-defined coordination environments of Bi, Al, and Cl. This simplicity leads to a clearer explanation of Bi related PL behaviors, and makes $Bi_5(AlCl_4)_3$ be a rather good model for
55 the study of NIR emitting materials containing Bi, although much effort is needed to synthesize high-quality single crystals. This work clearly revealed that $Bi_5^{3+}$ polycation, the average oxidation state of which is + 0.6, is one kind of optically active center. It is believed that detailed investigations of PL features from this type
60 of materials through structure and property characterizations as well as quantum chemistry calculation would be helpful to yield unsurpassed insights into Bi related PL behaviors, because of the establishement of structure-property relationships.[6b]

Aiming at this, here we report the PL behaviors from the $Bi_8^{2+}$
65 and $Bi_5^{3+}$ polycations stabilized by $[AlCl_4]^-$. The obtained product was studied in details by powder X-ray diffraction (PXRD), excitation-emission matrix (EEM) and time-resolved PL measurements as well as quantum chemistry calculations. Our results clearly revealed that $Bi_8^{2+}$ exhibits NIR emission peaking
70 at around 1180 nm, while $Bi_5^{3+}$ shows emissions at longer wavelengths. Furthermore, the results obtained from quantum chemistry calculation of $Bi_8^{2+}$ polycation helps us assign some excitation bands, especially in the visible spectral range, to specific electronic transitions of bismuth polycations.

## Experimental details

### Sample preparation

The product was synthesized by a modified method reported by Corbett *et al.* and Beck *et al.*.[1b,17] $AlCl_3$ (Sigma-Aldrich, 99.99%),
80 $BiCl_3$ (Alfa Aesar, ultra dry, 99.997%), Bi powder (Alfa Aesar, 99.999%) were used as received. Because of the high moisture sensitivity of the anhydrous metal halides used in this work, all manipulations were performed under dry $N_2$ atmosphere in a glovebox (< 2 ppm $H_2O$; < 0.1 ppm $O_2$). A mixture of $BiCl_3$, $AlCl_3$, and Bi was enclosed in a glass tube under vacuum and
85 heated in a muffle furnace.[1b] The mixture was first kept at 350 °C

for one week, then undergoes an initial cooling from 350°C to 160°C at a rate of 10 °C/h, and finally is cooled very slowly with 2°C/h from 160°C to 130°C. The obtained black product was seperated and sealed in bottles or capillaries for the following measurements.

**Materials characterization**

The product was characterized by X-ray diffractometer (Rigaku-RINT Ultima3, λ=1.54056 Å). The sample was grounded into powders, sealed in 1 mm Hilgenberg borosilicate capillaries and kept spinning during the measurement. The EEM was taken by Horiba NanoLog spectrofluorometer equipped with a monochromated Xe lamp and a liquid $N_2$ cooled photomultiplier tube (Hamamatsu, R5509-72). The powder was sandwiched between a glass microscope slide and a quartz cover glass and sealed with glue. Emission spectra were taken at different excitation wavelengths from 250 to 910 nm with 20 nm intervals. All spectra were corrected for spectral response of the detection system. Time-resolved PL measurements were performed by detecting the modulated luminescence signal with a photomultiplier tube (Hamamatsu, R5509-72), and then analyzing the signal with a photon-counting multichannel scaler. The excitation source for the time-resolved PL measurements was 600 nm light (pulse width: 5nsec; frequency: 20Hz) from an optical parametric oscillator pumped by the third harmonic of a Nd:YAG laser. The detailed quantum chemistry calculation on $Bi_8^{2+}$ polycation is shown in the following section.

**Results and discussion**

The crystallinity of as-prepared products was examined by PXRD (Fig. 1a). The diffraction pattern corresponds well to the simulated diffractogram determined from a $Bi_8(AlCl_4)_2$ single crystal (CSD no: 405159), as reported by Beck et al..[1b] In addition, a weak peak located at 10.41 ° is observed, which can be attributed to the (012) reflection of the $Bi_5(AlCl_4)_3$ phase (CSD no: 420082). It was reported that amorphous red phase usually forms when synthesizing $Bi_8(AlCl_4)_2$ crystal.[1b] However, detailed microscopic observation revealed that the product is dark black, and lacks red phase. This suggests that the ratio of $Bi_5(AlCl_4)_3$ to $Bi_8(AlCl_4)_2$ is rather low. The $Bi_8(AlCl_4)_2$ phase consists of $Bi_8^{2+}$ polycations and tetrahedral $[AlCl_4]^-$ anions (Fig. 1b). The $[AlCl_4]^-$ ions render the compounds air- and moisture-sensitive because of the hydrolytic instability of Al-Cl bond.

To examine the photophysical property, we took fluorescence EEM of the sample. As shown in Fig. 2, four peaks appear at the excitation/emission wavelengths of 290/1180, 490/1180, 636/1187, and 830/1192 nm, among which the intensity of the fourth peak is the strongest. Clearly, the PL peak does not display significant excitation-wavelength-dependent shift, which situate in the range of 1180-1200 nm under diverse excitation wavelengths (Figs.2 and 3a). However, the emission lineshapes are strongly influenced by the excitation wavelengths. Further detailed examination of the emission spectra reveals that the emission at 590 nm excitation is much more symmetrical than those at other wavelengths; notable emssion tails develop at longer wavelengths under the excitation of 290, 490 and 830 nm.

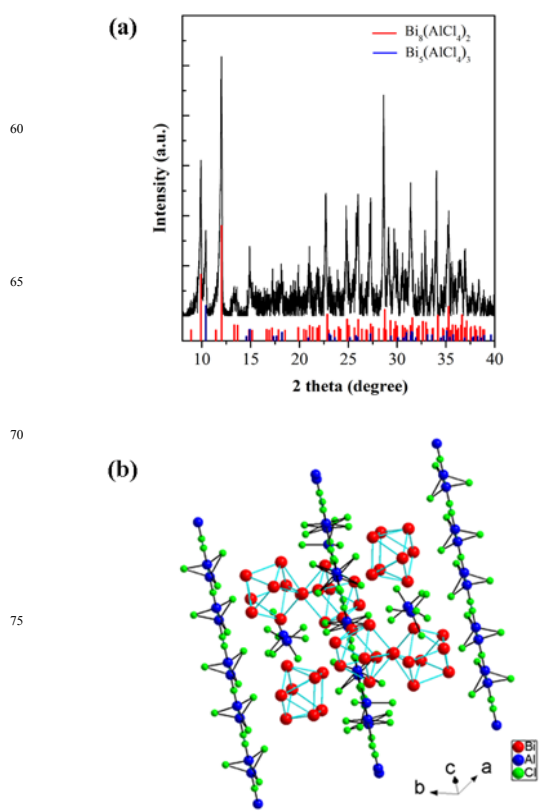

**Fig. 1** (a) PXRD of the synthesized product after removal of the background. The vertical red and blue lines represent the diffraction peaks of $Bi_8(AlCl_4)_2$ (CSD no: 405159) and $Bi_5(AlCl_4)_3$ (CSD no: 420082) phases, respectively. (b) General view of the structure of $Bi_8(AlCl_4)_2$

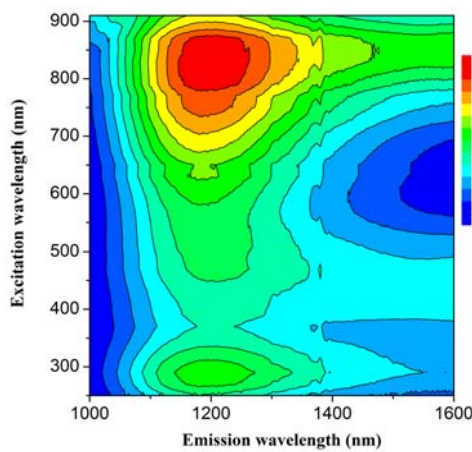

**Fig. 2** Contour EEM plot of the obtained sample. Note that the jump at around 1400 nm results from measurement artefacts.

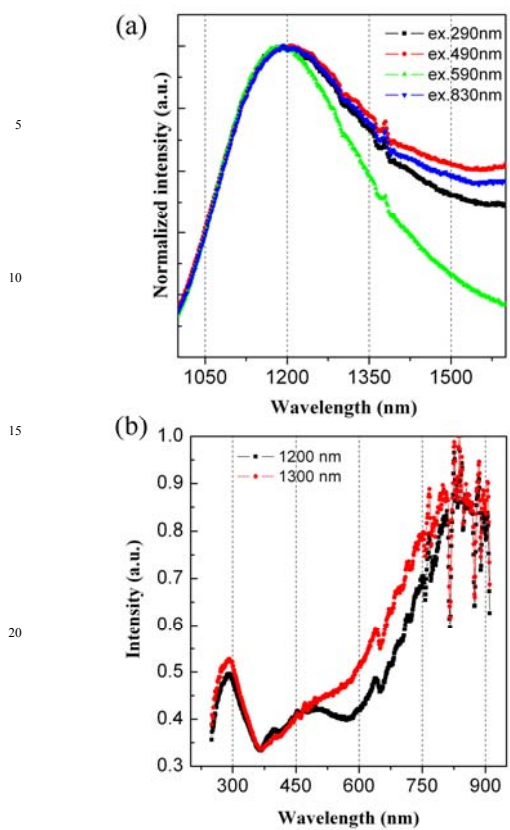

**Fig. 3** (a) NIR PL spectra from the sample under the excitation of 290, 490, 590 and 830 nm. (b) PL excitation (PLE) spectra of the sample when monitored at 1200 and 1300 nm. Note that the sharp lines between 750 and 900 nm are due to measurement artefacts. Owing to the weak emission from this sample, large slit widths for the excitation and emission were used to take the spectra.

The peculiar PL characteristics shown in Figs. 2 and 3 allow us to identify the imporant features of the obtained sample. Firstly, given that the down-converted emission from this sample strongly depends on the excitation wavelength, there should remain more than one optically active center in this system, all of which contribute to the broad NIR emission. Furthermore, the main emitter of this system contributes to the emssion peaking at ca. 1180 nm, while others to longer wavelengths. That is, simutaneous existence of multi-type active centers results in such broad and asymmetric emission profiles. Secondly, structural analysis suggests that the sample consists of $Bi_8(AlCl_4)_2$ phase accompanied by an impurity of $Bi_5(AlCl_4)_3$. Thus, it is reasonable to expect that the dominant emitter in this system is $Bi_8^{2+}$ polycation, while $Bi_5^{3+}$ polycation affects its emission behaviors as a result of overlapping excitation/emission bands between them. Interestingly, we observed that the emission tails from the present sample appear under the excitation of certain wavelengths (Fig. 3a). Our recent results revealed that high-quality molecular crystals containing $Bi_5^{3+}$ stabilized by $[AlCl_4]^-$ or $[GaCl_4]^-$ anions demonstrate rather broad NIR emission with an emission maxima > 1600 nm.[18a] Cao et al. reported that the emission from $Bi_5^{3+}$ can even extend over a wider range with a peak at around 1700 nm, although the final product is mixed with $NaAlCl_4$.[18b] All these facts lead us to infer that the tails stem from the electronic transition of $Bi_5^{3+}$ in $Bi_5(AlCl_4)_3$. Further comparison of the absorption and excitation bands of $Bi_8^{2+}$ and $Bi_5^{3+}$ suggests that an indirect excitation process for $Bi_5^{3+}$ should occur when exicited at 590 nm owing to its inherent weak absorption,[1b,17,18] which results in the weak emission at longer wavelengths. Thirdly, it is noteworthy that the absorption spectra of $Bi_8^{2+}$ and $Bi_5^{3+}$ polycations overlap in certain UV-Vis-NIR ranges,[14,17] however, both of which have their characteristic absorption bands. Therefore, we can use such kind of unique "fingerprint" to discriminate their distribution in the sample and contribution to the PL. For instance, $Bi_5^{3+}$ has a narrow absorption band in the range of 620-670 nm,[14] which agrees well with the observed PL and PLE behaviors (Fig. 2 and 3b). Furthermore, it is known that $Bi_5^{3+}$ demonstrates two separate main excitation bands peaking at around 513 and 815 nm, both of which show comparable intensity when $Bi_5^{3+}$ exists in high-quality $Bi_5(AlCl_4)_3$ crystal.[18] However, the sample shown here displays two main bands/shoulders at 490 and 830 nm in the range of 360-900 nm (Fig. 3b), the intensity of which shows a large difference. After thorough comparison of these PLE results with the diffuse reflectance spectrum of $Bi_8(AlCl_4)_2$,[17] it is reasonable to assign these two bands to the electronic transitions of $Bi_8^{2+}$. Fourthly, in addition to above excitation bands, an excitation band peaking at 290 nm appears, which was also found in the $Bi_5(GaCl_4)_3$ synthesized through the oxidation of Bi metal by gallium chloride salt, and can not be assigned to the electronic transition of $Bi_5^{3+}$ or $Bi_8^{2+}$.[18a] It is well known that $Bi^{3+}$ ion has an absorption band at around 290 nm corresponding to $^1S_0 \rightarrow {}^3P_1$ transition.[19] Thus, one possibility is that there remains unreacted $Bi^{3+}$ embedded in the product, which could act as an sensitizer for $Bi_5^{3+}$ and $Bi_8^{2+}$ polycations. However, now we can not rule out the contribution of structural defects or other impurities to this band, since it extends from 250 to 367 nm, much broader than that of $Bi^{3+}$.[19]

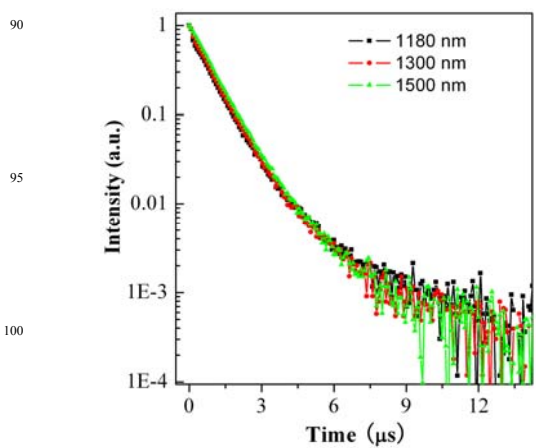

**Fig. 4** Transient emission decay profiles for the sample monitored at 1180, 1300, and 1500 nm under the excitation of 600 nm.

The above assignments were further reinforced by the result of time-resolved PL spectra. As shown in Fig. 4, the sample demonstrates nonexponential decays at 1180, 1300 and 1500 nm under the exciation of 600 nm pulsed light, evidencing that more than one active center gives rise to the emission. Furthermore, it is interesting to note that the emission at a longer wavelength appears a slower decay rate. Recent result revealed that the 1/e lifetime of NIR emission from $Bi_5^{3+}$ is ~ 4 μs.[14] Thus, it is believed that the slightly increased lifetime at longer emission wavelength is due to more contribution of $Bi_5^{3+}$ to the PL.

To obtain a rational explanation of the observed photophysical behaviors, next we performed a detailed quantum chemistry calculation on $Bi_8^{2+}$ polycation using the Amsterdam Density Functional (ADF) program package developed by Baerends *et al.*.[20,21] The geometry of $Bi_8^{2+}$ obtained from the XRD analysis was used for the following calculations.[1b] The detailed structural information was shown in Fig. 5. Spin-restricted density functional theory (DFT) was employed to determine energies and compositions of excited states of $Bi_8^{2+}$ polycation, using the Hartree-Fock method. The Slater type all-electron basis set utilized in the DFT calculation is of triple-zeta polarized (TZP) quality. The allowed and forbidden electronic transitions in the UV-Vis-NIR region were calculated using a Davidson method. The ADF numerical integration parameter was set to 3.0, and scalar relativistic effect was accounted for the DFT calculation.

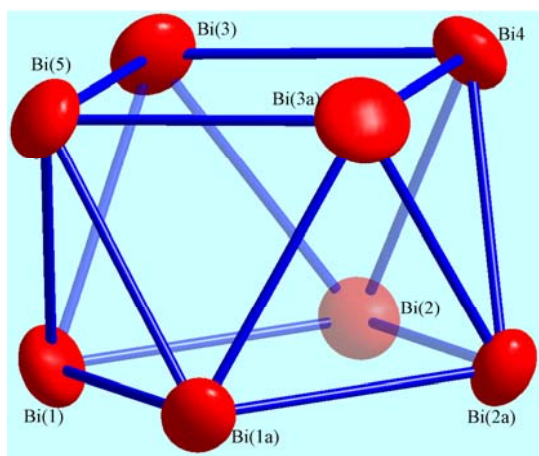

**Fig. 5** The antiprismatic $Bi_8^{2+}$ cation in Bi8(AlCl4)2. Selected bond length (Å): Bi(l)-Bi(2) 3.087(7), Bi(l)-Bi(3) 3.124(5), Bi(l)-Bi(5) 3.122(5), Bi(1)-Bi(1a) 3.078(9), Bi(2)-Bi(3) 3.093(5), Bi(2)-Bi(4) 3.105(5), Bi(2)-Bi(2a) 3.099(9), Bi(3)-Bi(4) 3.086(7), Bi(3)-Bi(5) 3.086(7). Selected bond angles (degree): Bi(1a)-Bi(l)-Bi(2) 90.2(1), Bi(1)-Bi(2)-Bi(3) 60.7(1), Bi(l)-Bi(3)-Bi(4) l02.7(1).

As shown in Table 1, $Bi_8^{2+}$ polycation should display strong visible electronic transitions in the range of 400-460 nm, owing to its allowed singlet-singlet excitation characteristics. In addition, this cation has forbidden-like transitions at longer wavelengths. Totally, we obtain 20 singlet-singlet and singlet-triplet excitation bands. We compare this theoretical result with experimental one shown in Figs. 2 and 3. Interestingly, we can see that in the range of < 500 nm, the experimental values are similar to theoretically determined ones, suggesting that these excitations can be ascribed to transitions No. 11-20 (Table 1). However, at longer wavelengths, there remains great discrepancy between experimental and calculated values, which was also observed when using other quantum chemistry softwares.[22] It is believed that this unsatisfactory consistence might result from the calculation method used here. Because of high atomic number of bismuth (Z=83), we should expect that spin-orbit coupling has significant effects on orbital energies and allowed/forbidden transitions. Indeed, the effect of strong spin-orbit coupling is possible to make singlet-triplet excitations accessible (i.e., the transitions become allowed). However, it is worth to note that taking spin-orbit coupling rather than scalar relativistic effect into consideration, when using ADF software for the excitation energies calculation, results in imaginary eigenvalue. It is believed that the selection of other quantum chemistry softwares may be helpful to further calculate the NIR theoretical excitation energies of $Bi_8^{2+}$.

**Table 1.** The calculated electronic transitions of $Bi_8^{2+}$ polycation. S-T and S-S represent the singlet-singlet and singlet-triplet transitions, respectively.

| No | Wavelength (nm) | Oscillator strength | Symmetry | Type |
|----|-----------------|---------------------|----------|------|
| 1  | 877.6 | 0 | A′ | S-T |
| 2  | 864.8 | 0 | A″ | S-T |
| 3  | 677.0 | 0 | A″ | S-T |
| 4  | 663.6 | 0 | A′ | S-T |
| 5  | 656.0 | 0 | A″ | S-T |
| 6  | 516.7 | 0 | A′ | S-T |
| 7  | 513.8 | 0 | A″ | S-T |
| 8  | 510.6 | 0 | A′ | S-T |
| 9  | 507.3 | 0 | A′ | S-T |
| 10 | 505.3 | 0 | A″ | S-T |
| 11 | 461.0 | $7.23 \times 10^{-6}$ | A″ | S-S |
| 12 | 442.8 | $4.82 \times 10^{-4}$ | A″ | S-S |
| 13 | 440.0 | $5.69 \times 10^{-4}$ | A′ | S-S |
| 14 | 417.5 | $3.71 \times 10^{-3}$ | A' | S-S |
| 15 | 416.5 | $4.11 \times 10^{-3}$ | A″ | S-S |
| 16 | 411.2 | $3.93 \times 10^{-3}$ | A′ | S-S |
| 17 | 408.4 | $9.84 \times 10^{-4}$ | A′ | S-S |
| 18 | 408.1 | $1.78 \times 10^{-4}$ | A″ | S-S |
| 19 | 401.6 | $7.92 \times 10^{-5}$ | A″ | S-S |
| 20 | 401.5 | $6.34 \times 10^{-5}$ | A′ | S-S |

Finally, we would like to comment the impact of the reported findings regarding the potential applications of such kinds of systems and the practical implications for solving the confusions associated with NIR PL in bismuth doped materials. First, a wide range of Bi polycations have been successfully synthesized during past decades.[1,12,14] However, little attention has been paid on the photophysical properties and the practical applications of these peculiar structures.[12,14,18] Systematical investigation of these systems is hopeful to find novel optically active centers that might meet the urgent needs in photonic areas. Second, given the structural simplicity of these molecular crystals, their photophysical characteristics can be used as smart fingerprints for the study of more complex systems containing Bi, such as glasses and conventional crystals. This work will certainly stimulate new experimental and theoretical treatments of Bi polycations to deepen the understanding of Bi related photophysical behaviors.

## Conclusions

In summary, the photophysical properties of $Bi_8^{2+}$ and $Bi_5^{3+}$ polycations stabilized by $[AlCl_4]^-$ units have been studied in details. The combined experimental evidence of XRD and PL results, taken in conjunction with the result of quantum chemistry calculation, makes it clear that both $Bi_8^{2+}$ and $Bi_5^{3+}$ contribute to the observed NIR emission. Furthermore, it was revealed that $Bi_8^{2+}$ polycation where Bi has an average oxidation state of + 0.25 results in the emission peaking at around 1180 nm, while $Bi_5^{3+}$ the longer-wavelength emission. It is clear that systematical examination of structural and luminescent properties as well as detailed quantum chemistry calculation of molecular crystals containing such kinds of bismuth units allows us to obtain a deeper understanding on bismuth-related photophysical behaviors, because of the establishment of structure-property relationships.[6b,12,14,18] Our work shown here not only enriches the species of luminescent polycations, but also serves to solve the confusions on the PL origin of Bi in other material systems such as glasses and conventional crystals.

## Acknowledgements

H. Sun gratefully acknowledges the funding support from the International Center for Young Scientists, National Institute for Materials Science, Japan (grant no. 215114). H. Sun greatly thanks the support from Prof. M. Fujii and Mr. Z. H. Bai in Kobe University for using lifetime measurement setup.

## Notes and references

*a International Center for Young Scientists (ICYS), National Institute for Material Sciences (NIMS), 1-2-1 Sengen, Tsukuba-city, Ibaraki 305-0047, Japan. Fax: +81-11-706-7881; E-mail: timothyhsun@gmail.com*
*b Division of Materials Science and Engineering, Faculty of Engineering, Hokkaido University, Sapporo 060-8628, Japan*
*c Advanced Ceramics Group, Advanced Materials Processing Unit, National Institute for Materials Science (NIMS), 1-2-1 Sengen, Tsukuba-city, Ibaraki 305-0047, Japan*
*d PRESTO, Japan Science and Technology Agency (JST), 4-1-8 Honcho Kawaguchi, Saitama 332-0012, Japan*